\documentclass[runningheads,a4paper]{llncs}
\usepackage{amssymb}
\usepackage{graphicx}
\usepackage{mathptmx,mathrsfs,amsmath}
\usepackage[squaren]{SIunits}
\usepackage{url}
\usepackage{subfig}
\urldef{\mailsa}\path|{sreenikg, pkumari, praseedha, sc}@ee.iitb.ac.in|
\urldef{\mailsb}\path|anna.kramer, leonie.kunz, christine.reiss, nicole.sator,|
\urldef{\mailsc}\path|erika.siebert-cole, peter.strasser, lncs}@springer.com|    
\newcommand{\keywords}[1]{\par\addvspace\baselineskip
\noindent\keywordname\enspace\ignorespaces#1}

\begin{document}
\mainmatter  % start of an individual contribution

\title{Haptic Rendering of Cultural Heritage Objects at Different Scales}

\titlerunning{Haptic Rendering of Cultural Heritage Objects}

\author{Sreeni~K.~G.%
% \thanks{Sreeni~K.~G. is with the Department
% of Electrical Engineering, Indian Institute of Technology, Bombay,
% India, e-mail: sreenikg@ee.iitb.ac.in.}%
% \and Priyadarshini~Kumari 
\and Priyadarshini~K. 
% \thanks{Priyadarshini~Kumari is with the Department
% of Electrical Engineering, Indian Institute of Technology, Bombay,
% India, e-mail: pkumari@ee.iitb.ac.in.}
\and Praseedha A.~K. 
% \thanks{A.~Praseedha~Krishnan is with the Department
% of Electrical Engineering, Indian Institute of Technology, Bombay,
% India, e-mail: praseedha@ee.iitb.ac.in.}
\and Subhasis~Chaudhuri
\thanks{This work was supported in part by a DST grant on Indian Digital Heritage and another by MCIT on perception engineering.}
}
% \thanks{This work was supported in part of a DST grant on Indian digital heritage and another by MCIT on perception engineering.}
\authorrunning{Haptic Rendering of Cultural Heritage Objects}


 \institute{Vision and Image Processing Laboratory,
        Department of Electrical Engineering,\\~Indian Institute of Technology Bombay, Powai, Mumbai-400076\\
\mailsa\\
% \mailsb\\
% \mailsc\\
% \url{http://www.springer.com/lncs}
}


\toctitle{Lecture Notes in Computer Science}
% \tocauthor{Authors' Instructions}
\maketitle
\begin{abstract}
In this work, we address the issue of virtual representation of objects of cultural heritage for haptic 
interaction. Our main focus is to provide a haptic access of artistic objects of any physical scale to 
the differently abled people. This is a low-cost system and, in conjunction with a stereoscopic visual 
display, gives a better immersive experience even to the sighted persons. To achieve this, we propose a 
simple multilevel, proxy-based hapto-visual rendering technique for point cloud data which includes the 
much desired scalability feature which enables the users to change the scale of the objects adaptively 
during the haptic interaction. For the  proposed haptic rendering technique the proxy updation loop runs 
at a rate 100 times faster than the required haptic updation frequency of 1KHz. We observe that this 
functionality augments very well to the realism of the experience. 
\keywords{Haptic rendering, HIP, proxy-based rendering, voxel based rendering, image pyramid, virtual museum, stereoscopic display.}
\end{abstract}
\section{Introduction}\label{liter}
In the recent years digital technology is 
paving a way into safeguarding cultural heritages, and it also offers a great promise for enhancing 
access to them. A user's experience of accessing such cultural objects can be made more realistic 
and immersive by incorporating the recently evolving haptic technologies. Museum of Pure Form~\cite{Berg02}, 
a virtual reality system placed inside several museums and art galleries around Europe is an attempt 
to use of haptic technologies in cultural heritage applications. The incorporation of haptics in cultural 
heritage applications also helps 
in letting visually impaired people feel the exhibits that are behind glass enclosures, making 
even very fragile objects available to the scholars and allowing museums to show off a range of 
artefacts that are currently in storage due to lack of space. Further a joint hapto-visual rendering 
improves the immersivenes of the kinesthetic interaction. Some existing systems also allow users 
to interact with museum exhibition pieces via the internet~\cite{Zhan01}. It is required that such a system should enable the users to
hapto-visually explore ancient monuments and heritage sites like Taj Mahal. However, currently available haptic systems are unable to 
handle objects at different scales.

As a part of our exercise in preserving our cultural heritage we propose a simple multilevel 
hapto-visual rendering technique with depth data of cultural heritage objects.
With mesh models of objects there are effective rendering techniques in haptics like god object rendering 
algorithm as proposed in~\cite{Zill95}. However this algorithm fails in the case of point cloud 
based models. Further cultural objects appear at various different scales, and the user needs to 
experience the object at different levels of details. A mesh-based haptic technique is not amenable 
to scale changes as it requires the mesh to be pre-computed at all scales which is not feasible.
In this paper, we propose a fast, proxy based rendering technique capable of working with point 
cloud based $3-D$ models. Additionally the proposed method is amenable to haptic rendering at 
various scales. In order to render the model at different levels of resolution, we generate 
depth at each point of the model by reading the contents of depthbuffer in OpenGL and create a 
Monge surface from it. We show that the user's experience can be improved by allowing the user 
to interact with the object at multiple resolutions. This feature allows the user to feel the 
object more precisely at a closer level when needed and zoom out when context is desired. We have also developed a 
graphical user interface to make accessibility easier. Moreover, the easy availability of 
$3-D$ models makes it a cost-effective system to savour the experience of various cultural 
heritage sites. The key contribution in this paper include how to render a Monge surface represented by a 
non-uniform point cloud data and how to handle scale change for zooming in 
and out during haptic interaction.
\section{Literature Review}\label{lit_sec}
In the haptic rendering literature 
there are mainly two different approaches: Polygon (geometry) based rendering and Voxel based 
rendering. A good introduction to the basic haptic rendering technique is given by 
~\cite{Lay07}, \cite{Sali04}.
Traditional haptic rendering method is based on a geometric surface representation which 
consists of mainly triangular or polygonal meshes. In polygon based rendering, each time the haptic 
interface point (HIP) penetrates the object, the haptic rendering algorithm calculates the 
closest surface point on the polygonal mesh and the corresponding penetration depth. If $\mathbf{d}$ 
is the vector representing the depth of penetration in the model, the reaction force can be calculated as $\mathbf{F}=-k\mathbf{d}$, 
where $k$ is the stiffness constant, a physical property of the associated surface.
The above method has problems while determining the appropriate direction of the force while 
rendering thin objects. Zilles and Salisbury~\cite{Zill95}, and Ruspini \emph{et al.}~\cite{Rusp97} 
independently introduced the concept of god-object and proxy algorithm, respectively, which 
can solve the problems associated with thin objects. 

In the God-Object rendering method ~\cite{Zill95}, the 
authors use a second point in addition to the HIP called ``god-object", sometimes called the 
ideal haptic interface point (IHIP). While moving in free space the god-object and the HIP 
are collocated. However, as the HIP penetrates the virtual object, the god-object is constrained 
to lie on the surface of the virtual object. The position of the god-object can be determined 
by minimizing the energy of the spring between the god-object and the HIP taking into account 
constraints represented by the faces of the virtual object~\cite{Lay07}. If $(x,y,z)$ are the 
coordinates of the proxy lying on the virtual object and ($x_h, y_h,z_h$) represents the 
coordinates of the HIP, the spring energy is given by
\begin{equation}\label{eq1}
L=\dfrac{\mathbf({x-x_h})^2}{2}+\dfrac{\mathbf({y-y_h})^2}{2}+\dfrac{\mathbf({z-z_h})^2}{2}+\sum\limits_{i=1}^{3}l_i(A_ix+B_iy+C_iz-D_i)
\end{equation}
where $L$ is the cost function to be minimized, $l_1$,  $l_2$,  $l_3$ are Lagrange multipliers 
and ($A_i$,  $B_i$,  $C_i$, $D_i$) are the homogeneous coefficients for the constraint plane 
equations on which the proxy lies. The `force shading' technique (haptic equivalent of Phong shading) 
introduced by Morgenbesser and Srinivasan refined the above algorithm while rendering smooth 
objects~\cite{Morg96}. 
% One common problem with the mesh based representation is that when the 
% object is not fully enclosed by the bounding planes, a small hole may remain and therefore the 
% IHIP sinks during the rendering process. 
Mesh based haptic rendering is not amenable to object scaling as 
the constraint equation for the planes ($A_i$,  $B_i$,  $C_i$, $D_i$) must be recomputed.

Volume haptic rendering technique is another alternative rendering technique used in haptics.
The most basic representation for a volume is the classic voxel array in which each discrete 
spatial location has a one-bit label indicating the presence or absence of material. Avila \emph{et al.} 
have used additional physical properties like stiffness, color and density 
during the voxel representation~\cite{Avil96}. The voxmap-point shell algorithm uses the voxel 
map for stationary objects and point shell for dynamic objects~\cite{Mcne99}, \cite{Renz01}. 
Point shell has been defined as a set of point samples and associated inward facing normals.  
However, these normals are not available and one needs to compute the normal at every location.
The external surface $\partial{O}$ of a solid object $O$ can be  described by the implicit 
equation as~\cite{Kim02}
\begin{equation*}
\partial{O} = \{{(x, y, z)\,\,\in \,\, R^3\,\, |\,\, \phi(x, y, z) = 0}\},
\end{equation*}
where $\phi$ is the implicit function (also called the potential function) and $(x, y, z)$ 
is the coordinate of a point in $3-D$. In other words, the set of points for which the potential 
is zero defines the implicit surface. This has found applications in haptic rendering. This 
technique also suffers from the thin object problem. 
% In  order to avoid such problems some 
% researchers have proposed explicit mesh reconstruction from the implicit surface. In the area 
% of surface  reconstruction there are several approaches such as Delaunay triangulation \cite{Koll04}, 
% alpha shapes \cite{Bern99} or Voronoi diagram \cite{Amen98}, \cite{Amen01}. These schemes typically 
% create a triangular (polygonal, in general) mesh that interpolates all or most of the points and 
% the associated surface normals help in the rendering process. However, these methods cannot deal 
% with point cloud data. 
Lee \emph{et al.} have proposed a rendering technique with point cloud data which computes the 
distance from HIP to the closest point on the moving least square (MLS) surface defined by the 
given point set\cite{Lee07}. Here the same problem occurs as in the distance field based rendering 
technique, since we do not keep track of HIP penetration and therefore is not good in rendering thin 
objects. El-Far \emph{et al.} used axis aligned bounding boxes to fill the voids in the point cloud 
and then rendered with a god object rendering technique\cite{Naim08}.
Leeper \emph{et al.} described a constraint based approach of rendering point cloud based data where 
the points are replaced by spheres or surface patches of approximate size \cite{Leep11}.
Another proxy based technique of rendering dense 3D point cloud was proposed in \cite{Sree12},
where the surface normal is estimated locally from the point cloud.
% We try to overcome these limitations by introducing the proposed method 
% explained in section \ref{sec:3}. Once the surface is reconstructed from the point cloud data 
% we can use the god object rendering technique given in equation (\ref{eq1}). \\
\section{Proposed Method}\label{meth_sec}
The rendering technique we propose is a proxy based method which does not use polygonal meshes 
for the reasons mentioned earlier. In practice, most of the cultural objects are carefully 
preserved and a dense $3-D$ scan is performed on these objects to create virtual $3-D$ model 
in the form of .obj, .ply, .3ds file, etc. Instead we directly use the point cloud data defining the models. As 
mentioned in the introduction, we get the depth data in the form  of $z_i=f(x_i,y_i)$ 
where $x_i$ and $y_i$ are discrete values and $z_i$ is the height of each sampled point 
from a reference plane ($z=0$) and $i$ and $j$ can take values depending on the size of the model. 
We haptically render the sampled surface of the object approximated by the depth values. 
In order to haptically render the object, we need to find the collision 
of HIP with the bounding surface and hence the penetration depth of HIP into the surface. 
The key factors in haptic rendering algorithm are-
{
\begin{enumerate}
\item The magnitude of the haptic force should be proportional to the penetration depth 
of HIP from the surface. %It is assumed in the paper that the material property (stiffness) of 
%the object does not change at different locations. If it does, this must be specified so that 
%an appropriate force may be rendered. 
\item The direction of force should be normal to the surface at the point of contact of the proxy.
\end{enumerate}
} 
By taking these factors into consideration, the proposed algorithm tries to move the proxy 
over the object surface in short steps during the interaction so that each time it finds 
the most appropriate proxy position, the new position minimizes the distance between HIP and proxy 
and at the same time applies the reaction force normal to the surface at the point of contact. 
Initially, let us assume that the surface is known or the values of $z$ is known for all values 
of $x$ and $y$ in continuum. 
\begin{figure}
\centering
\includegraphics[width=0.55\textwidth]{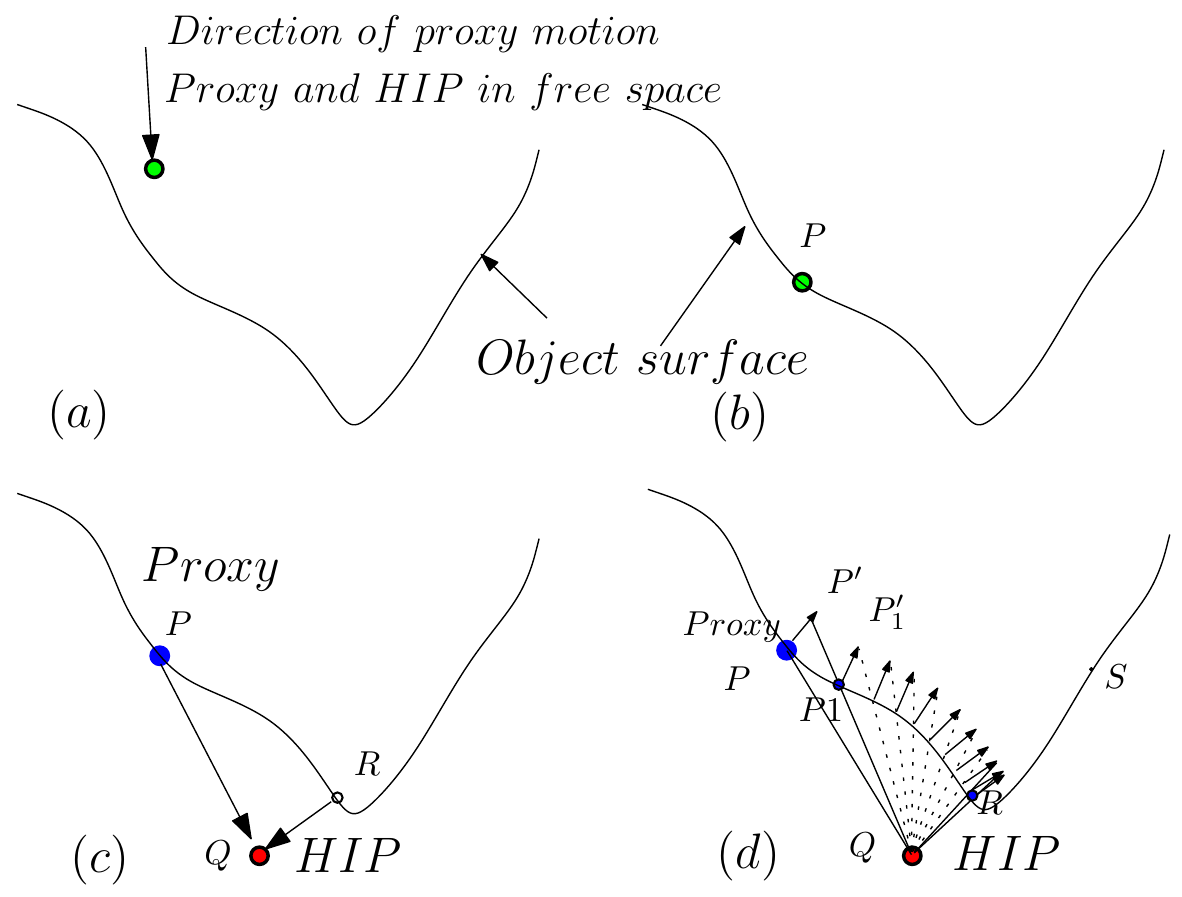}
\centering
\caption{Illustration of the proposed method to find the penetration depth of HIP into the surface.}
\label{mini_fig}
\end{figure} 

To understand our procedure let us look at the situation in Fig. \ref{mini_fig}. The bounding 
surface of the object is shown with the curve. In free space, proxy and HIP are collocated 
and is shown with green circle above the surface. Let the HIP and the proxy be in free space 
at a time $t=t_0$ as shown in Fig. \ref{mini_fig}(a). The HIP and hence the proxy are together 
moving towards the surface in a direction as shown by the arrow. At $t=t_1$, let HIP and proxy 
touch the surface at point $P$ as in Fig. \ref{mini_fig}(b). Up to this point the proxy moves 
with the HIP.  If the HIP is moved further in the direction it penetrates the surface and 
let $Q$ as shown in Fig. \ref{mini_fig}(c) be the HIP position at time $t=t_2$. Now the proposed 
algorithm finds the most appropriate position of proxy at $R$ where the distance $QR$ is minimum, 
and the penetration depth $QR$ is calculated in the direction exactly opposite to the surface 
normal at $R$.
To find the point $R$ from the starting point $P$ we  use the successive approximation method 
and move the proxy $P$ to a distance $\delta\mathbf{n}$ along the normal to the surface at point $P'$ 
and draw a line $P'Q$ to the current HIP as shown in Fig. \ref{mini_fig}(d). The point $P_1$ on 
the surface at which the line intersects is found and is updated as the current proxy position. 
Again we move the proxy point $P_1$ along the normal at the current proxy position to $P_1 '$ 
and the process is repeated until the final proxy position $R$ is attained. This is a greedy 
method but works well for smooth surfaces. If the surface has a fine texture, the line $P'Q$ 
may intersect the surface at multiple points and the process may converge to a poor, local 
minima, yielding a jerky haptic interaction. In case of multiple intersections, the one closest 
to $P'$ is selected. The length of the vector $\delta\mathbf{n}$ determines the rate of convergence of the process. 
A large value of $\delta\mathbf{n}$ may lead to some spurious interaction when $PP'$ may intersect the 
surface along the segment $RS$ instead of $PR$. Hence a smaller value of $\delta\mathbf{n}$ (0.1mm is used
in our algorithm) is preferred for rendering purposes. 
Quite naturally, it is required that the proxy position updation is 
performed within 1 ms of time, so that the user's interaction with the object through the haptic 
device is unhindered and is carried at 1 KHz.
\begin{figure}
\centering
\includegraphics[width=0.48\textwidth]{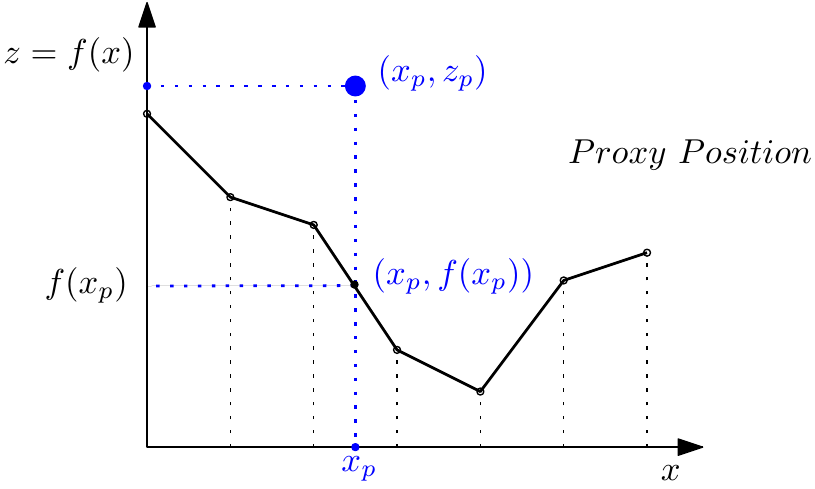}
\centering
\caption{Surface approximation from depth values.}
\label{bili_fig}
\end{figure} 
Till now, we have assumed the surface to be known. Now, we try to approximate the surface 
from the given depth values. In case of $2-D$ depth data, 
we project the proxy onto the X-Y plane and the corresponding depth value is obtained by 
interpolating the neighbourhood depth values to form a continuous function $z=f(x,y)$.
For better understanding, we consider a one-dimensional function $z=f(x)$ as shown 
in Fig. \ref{bili_fig}. 
As we have the function defined only at sampled points we 
interpolate the function at $x_p$ to find the value of $f(x_p)$. Since the available points 
are sampled quite densely, bilinear interpolation is sufficient to find the bounding surface 
as shown in Fig. \ref{bili_fig}.  
In order to check the collision of HIP with the function we perform the following. At a 
given proxy position $(x_p,z_p)$ we check for the function value $f(x_p)$. If $f(x_p)>z_p$ 
proxy has touched the surface, otherwise it is free to move towards the HIP. Extending the concept to $2-D$ depth data, 
let $\mathbf{X}_h=[x_h,y_h,z_h]^T$ denotes the HIP point and $\mathbf{X}_p=[x_p,y_p,z_p]^T$
denotes the proxy point. Collision can be easily checked here by comparing $z_p$ with the 
depth interpolated at the projected point $z=f(x,y)$. The proxy movement during the rendering is
managed by equation \ref{math_eq}. 
% The parameter $\alpha<1$ and it decides the speed of proxy 
% movement in free space and the value selected is 0.005 in our technique. 
\begin{eqnarray}\label{math_eq}
\mathbf{X}_p^{(k+1)} &=& \mathbf{X}_p^{(k)}+ \mathbf{\delta n}\hspace{0.8cm}  if \hspace{0.8cm} z_p<f(x_p,y_p) \nonumber\\ 
                     &=& \mathbf{X}_p^{(k)}+|\delta\mathbf{n}| \dfrac{(\mathbf{X}_h-\mathbf{X}_p)^{(k)}}
{|\mathbf{X}_h-\mathbf{X}_p|^{(k)}}\hspace{0.4cm} otherwise 
\end{eqnarray}
This allows a smooth interaction when the force is withdrawn out of the object. The updated proxy then 
slowly moves towards the new HIP position.
\section{Rendering}\label{sec:4}
\begin{figure}
\centering
\subfloat[]{\label{mesh_fig}\includegraphics[width=0.3\textwidth]{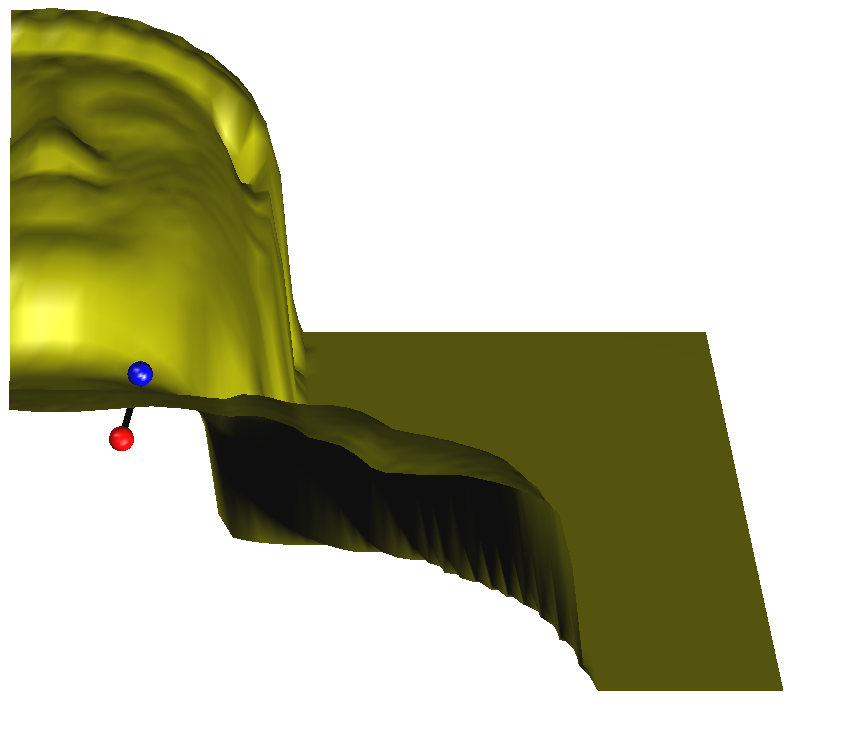}}         
\subfloat[]{\label{god_fig}\includegraphics[width=0.3\textwidth]{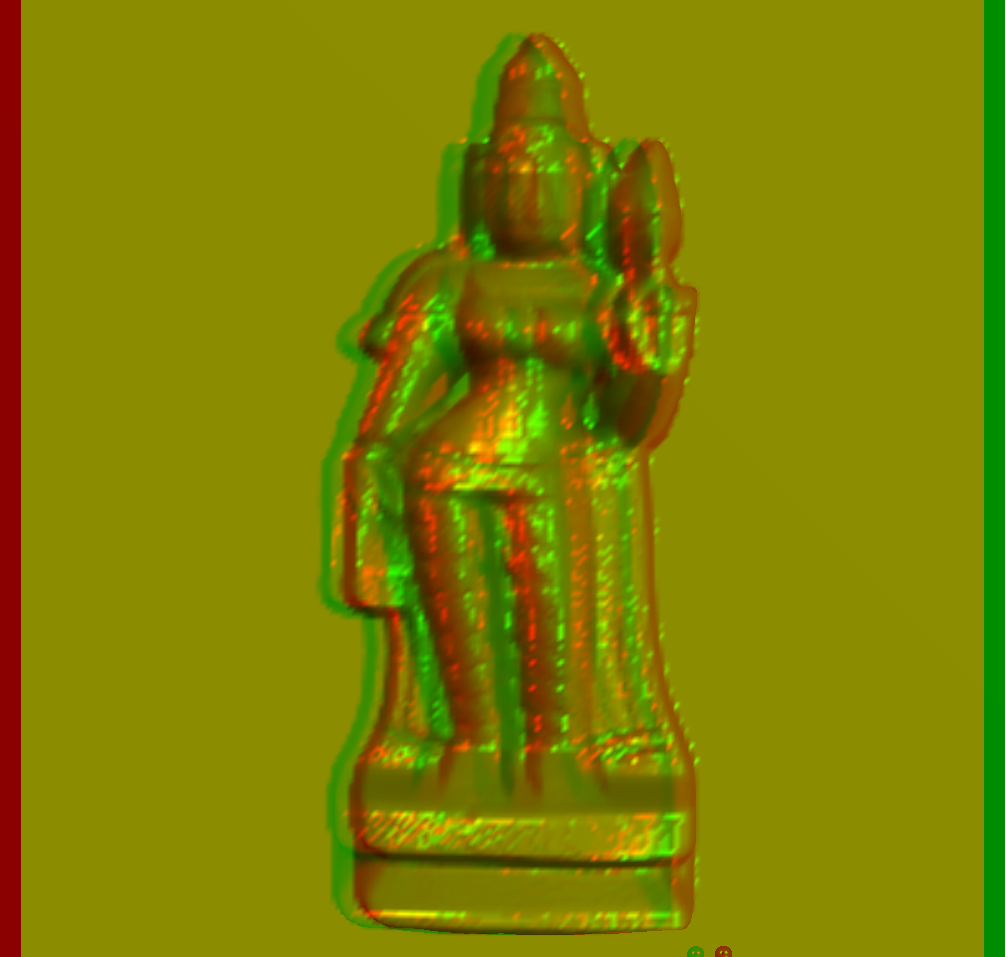}}
\centering
\caption{(a) Illustration of proxy and HIP positions for an arbitrary surface. (b) Stereoscopic view of 
an Indian heritage object. (Data Courtesy: \emph{www.archibaseplanet.com}) }
\end{figure}

Rendering part of our work concerns with both haptic rendering and graphic rendering. Haptic 
rendering involves generating software controlled forces and feeding it to the users to provide them
the sensation of touch. 
% Similarly, graphic rendering generates images of objects for visual display.
% \subsection{Haptic Rendering}
Any haptic rendering technique must include two steps:
{
\begin{enumerate}
\item detection of collision of the HIP with the object.
\item force computation if a collision is detected. 
\end{enumerate}
}
If $z_p<f(x_p,y_p)$ then the proxy has touched the 
object and a force needs to be fed back by the haptic device.
Subsequently, the reaction force is computed 
as $\mathbf{F}=-k\mathbf{d}$ where $k$ is the Hooke's constant, and $\mathbf{d}$ is the penetration 
depth given by $\mathbf{d}=|\mathbf{X}_h-\mathbf{X}_p|$,
where $\mathbf{X}_h$ is the HIP position and $\mathbf{X}_p$ is the proxy position. Here we assume the stiffness 
to be constant everywhere on the surface of the object, but it can also be a function of position, 
provided the material property of the object is well documented.
Fig. \ref{mesh_fig} shows proxy and HIP positions while rendering an arbitrary surface. For 
illustration we have selected only a small part of the depth map around the active region. 
The blue ball is the computed proxy touching the surface while the HIP is penetrated 
inside the surface. The HIP is shown with a red ball in the scene and the line from HIP to 
proxy is normal to the surface point at the proxy position.
In order to show the surface of the object graphically for simultaneous visual immersion, we 
display the image as a simple quad mesh out of the depth values. The normal is computed at 
each vertex. Although, we use point cloud data for haptic rendering, using the same for 
graphic rendering would result in gaps in the visually rendered object. Hence we have opted 
for the mesh-based graphical display in order to give a better perception to the viewer. 
We have used the stereoscopic display technique for creating the effect of depth in the 
image by presenting two offset images  in different colours separately to the left and 
right eye of the viewer. A human observer combines these $2-D$ offset images to recreate the $3-D$
perception. Anaglyphic glasses can be used to filter offset images from a single source, separated 
to each eye to give the perception of a  $3-D$ view to the users. Fig. \ref{god_fig} shows the 
$3-D$ view of an Indian heritage object as displayed on the screen.
% The mesh structure hence created is shown in Fig. \ref{fig:4}.
\subsection{Rendering at Different Scales}
\begin{figure}
\centering
\includegraphics[width=0.4\textwidth]{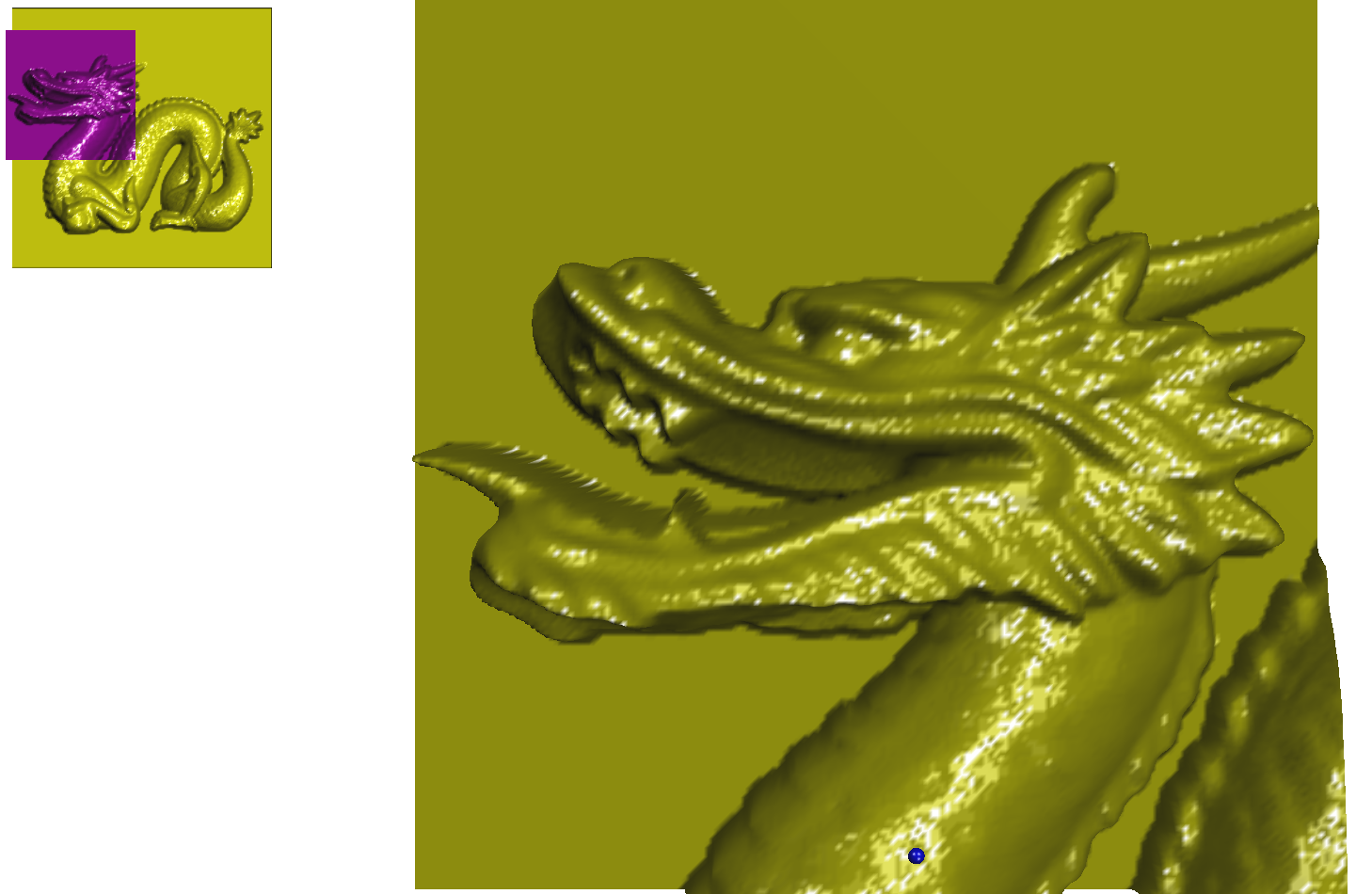}
\centering
\caption{Illustration of selection of a window for graphic and haptic rendering. 
(Data Courtesy:\emph{www.cc.gatech.edu/projects/large\_models}) }
\label{drag_fig}
\end{figure} 
As mentioned earlier, heritage objects come at various physical scales- a few cm$^2$ for 
coins and bas-reliefs to a several km$^2$ for ancient ruins like Hampi. In a virtual museum, 
one should be able to experience objects of all sizes at different scales to get a sense of 
overall structure to a finer details from the same data set. Hence, we have implemented adaptive 
scaling in both graphic and haptic domains. In order to scale the surface we resize depth data 
of resolution $N \times N$ depending on the level we want, with $N \times N$ as the lowest level. 
If we load the level $N \times N$ into the haptic space the full object can be rendered visually 
as well as haptically. Users can select the level as well as the region of interest at run time 
either using buttons in the haptic device or using keyboard functions. Additionally, we have developed a graphical user interface 
for easy acessibility. The pink window in Fig. \ref{drag_fig} represents the selected region to be zoomed in. 

Depending on the scale selected by the user, only the corresponding depth data is  dynamically 
loaded into the active haptic space and an appropriate haptic force is rendered. As only a limited 
subset of data is loaded, the rendering is very fast. In general, at higher levels of resolution, 
the user should be able to view higher depth value at each point and also more finer details. The 
haptic force also vary accordingly. Hence in order to incorporate realistic haptic and graphic 
perception, we need to appropriately scale the depth values at each level of depth map. Further, 
trying to map a large physical dimension over a small haptic work space (typically about 4 inch 
cube of active space) leads to a lot of unwanted vibrations (something similar in concept to aliasing) 
during rendering. Hence the depth values need to be smoothed before being downsampled and mapped into the haptic 
work space. The next section explains the generation of different levels of depth data. 
\section{Multi-Scale Data Generation}\label{sec:5}
\begin{figure} 
\centering
\includegraphics[width=0.65\textwidth]{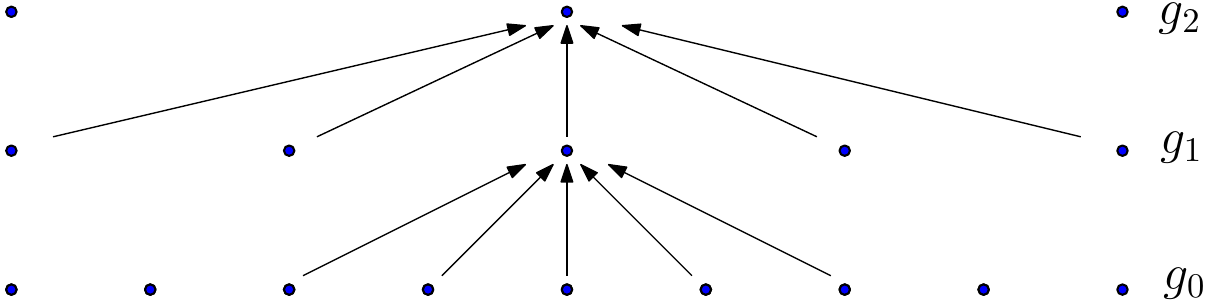}
\centering
\caption{A one-dimensional graphical representation of the process  which generates a Gaussian 
pyramid. Each row of dots  represents nodes within a level of the pyramid. The value of each 
node in the zero level is the magnitude of the corresponding depth map. The value of 
each node at a higher level is the decimated and weighted average of node values in the preceding level.}
\label{pira_fig}
\end{figure} 
The aim of this work is to allow users to have access to the cultural heritage at different 
levels of details. To obtain the depth data at different levels of details, we perform Gaussian 
low-pass filtering followed by down sampling with a factor $M$ where $M$ can be any integer. 
We can also use fractional values of $M$, but it requires rational function approximation 
methods. In our work, we illustrate with $M=2$. For that the data pyramid offers a flexible 
and convenient multiresolution format that mirrors the different levels of details~\cite{Gluc06}. 
It consists of the available highest resolution depth data and a series of successively lower 
resolution data. Low-pass filtering before 
sub-sampling is done to prevent  aliasing of data. Consequently, instability in the haptic 
domain is also prevented as smoothing removes higher frequency components responsible for micro 
textures on the surface. The presence of micro textures would have made sensing more realistic, 
but this makes the haptic rendering process miss to some extent a full understanding of the 
object at hand. The fine texture is experienced when the object is rendered at a finer scale 
by zooming into the object. The base, or zero level of the pyramid is equal to the original 
depth map ($g_0$). Level 1 of the pyramid corresponds to depth map $g_1$ which is reduced or 
low-pass filtered version of $g_0$. Each value in level 1 is computed as a weighted average 
of values in level 0 within a $5\times5$ window. Each value in level 2 ($g_2$) is then obtained 
from values of level 1 by  applying  the same pattern of  weights. Fig. \ref{pira_fig} shows the Gaussian
pyramid of depth map. The depth value at each point at the level $l$ is given by the following equation: 
\begin{equation}
g_l(i,j) = {\sum_{m=-2}^{2}\,\,\,\sum_{n=-2}^{2} w(m,n)g_{l-1}(2i+m,2j+n), }
\end{equation}
For levels  $1<l<N+1$, and nodes $(i, j)$, $0< i < C_l$, $0 < j < R_l$, the upper level of the 
pyramid can be represented in the above given form. Here $N$ refers to the number of levels in 
the pyramid, while $C_l$ and $R_l$  are the dimensions at level $l$. The weighting pattern 
$w(m,n)$ is the Gaussian kernel~\cite{Burt83}.

\section{Results}\label{sec:12}
The proposed method was implemented in visual C++ in a Windows XP platform with a CORE 
2QUAD CPU @ 2.66 GHZ with 2 GB RAM. We have experimented with various models of cultural heritage 
objects and a few of them are displayed below. The Fig. \ref{gane_fig} shows the model of Ganesh, 
visually rendered in OpenGL. For haptics rendering we use HAPI library. The blue ball represents the position of the 
proxy constrained to be on the surface. The discrete position in the model is displayed in 
a fixed $200\times200$ haptic space. The size and spatial resolution of the model depend on
two factors: the active space of the haptic device used to render the model, and the resolution 
at which the model should be displayed. We use a 3-DOF haptic device from NOVINT with a 4 inch 
cube of active space. While interacting with the object haptically, the average proxy updation time 
is 0.0056 ms which is much less than the required upper bound of 1 ms, and hence the 
user has very smooth haptic experience. The average time required for dynamic data generation 
and loading it into the haptic space depends on the resolution of input depth data and it was observed to be 
around 6.5 s and 2.0 s respectively for depth data with resolution of $800\times800$. As explained in the previous section, Fig. \ref{gane_figa} 
corresponds to the lowest level of details. We also carry out the rendering at finer 
levels of details by successively zooming into the heritage object. These are shown in Fig. \ref{gane_figb} and Fig. \ref{gane_figc}.
\begin{figure} 
  \centering
  \subfloat[]{\label{gane_figa}\includegraphics[width=0.33\textwidth]{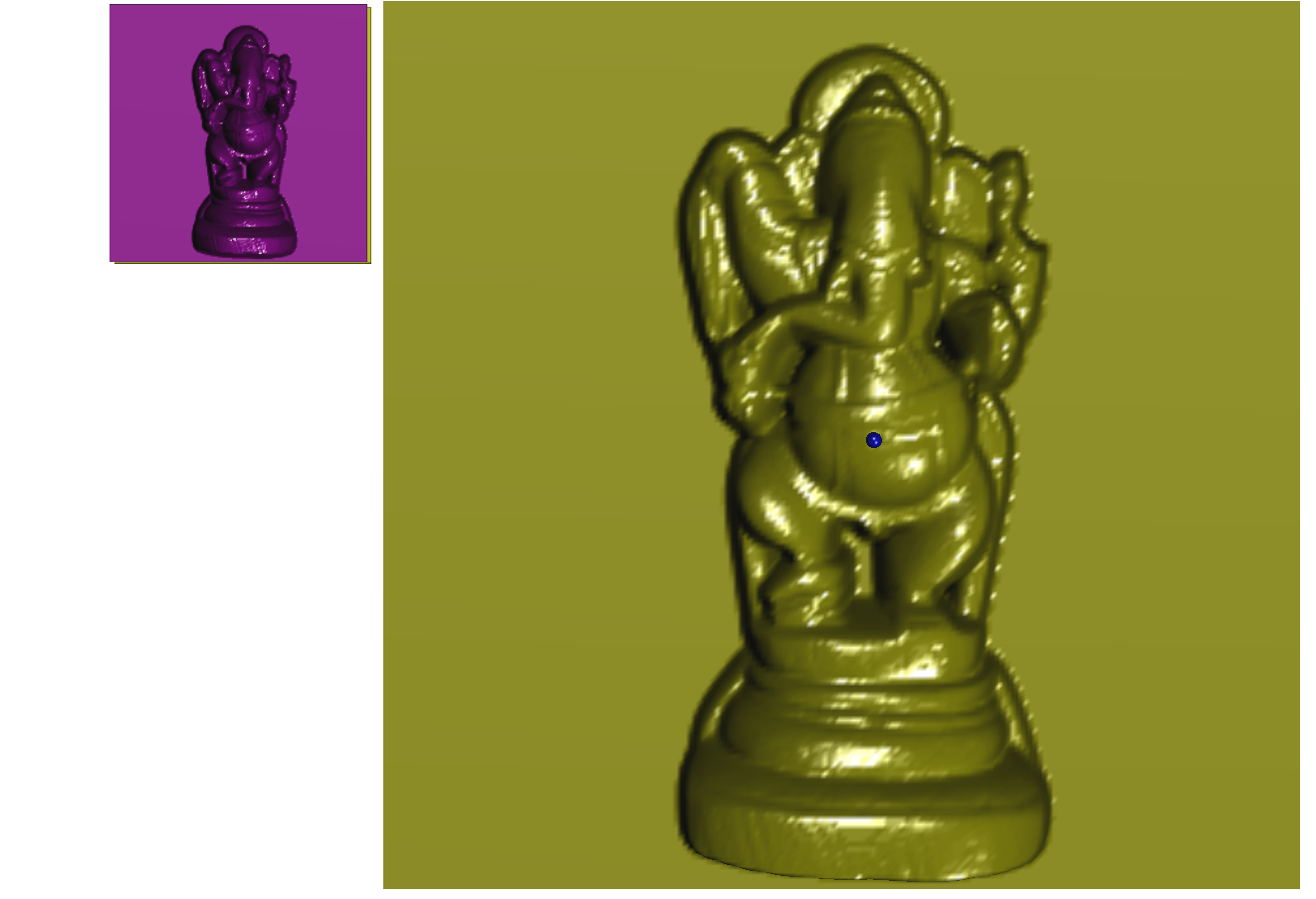}}         
  \subfloat[]{\label{gane_figb}\includegraphics[width=0.33\textwidth]{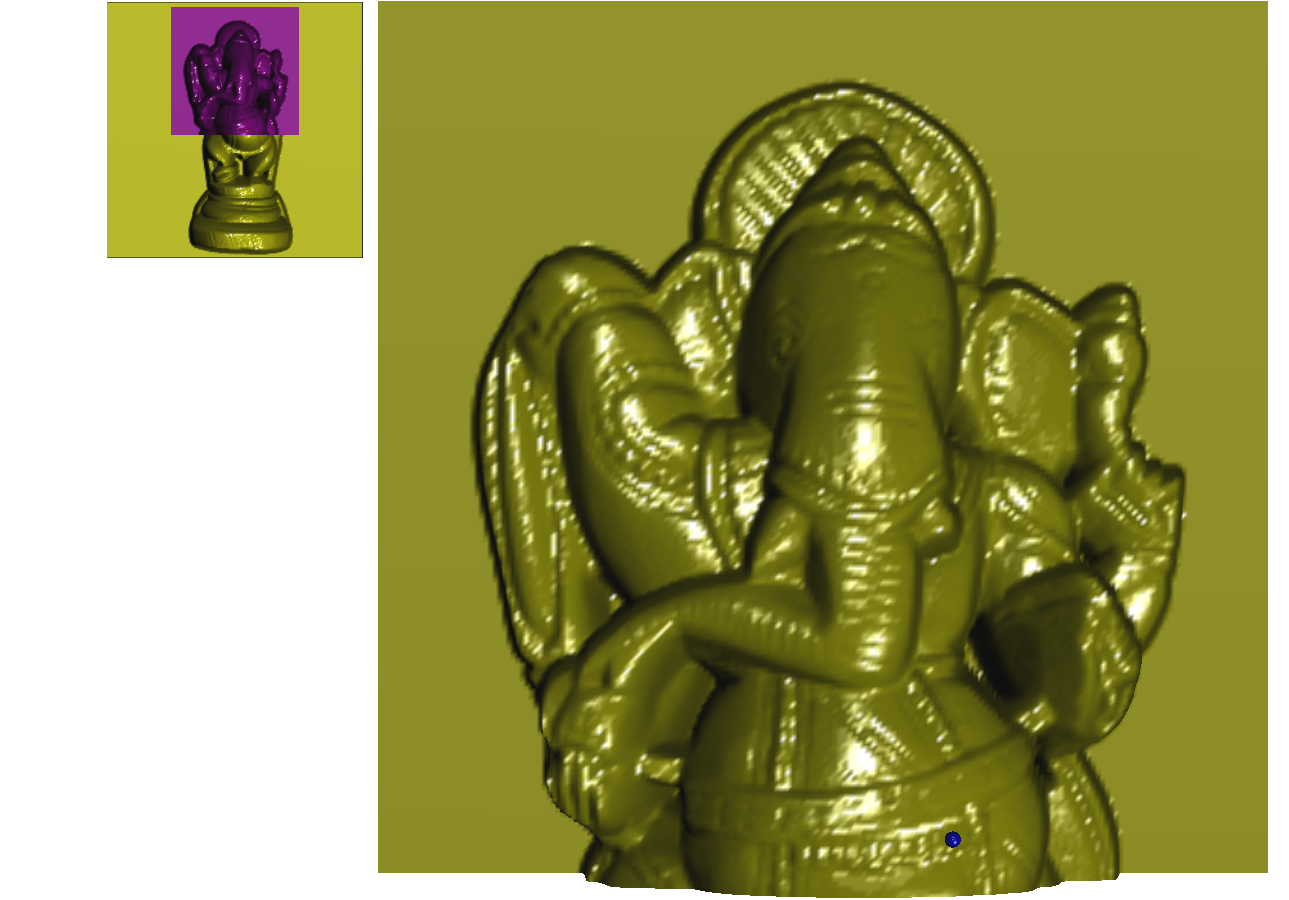}}
  \subfloat[]{\label{gane_figc}\includegraphics[width=0.33\textwidth]{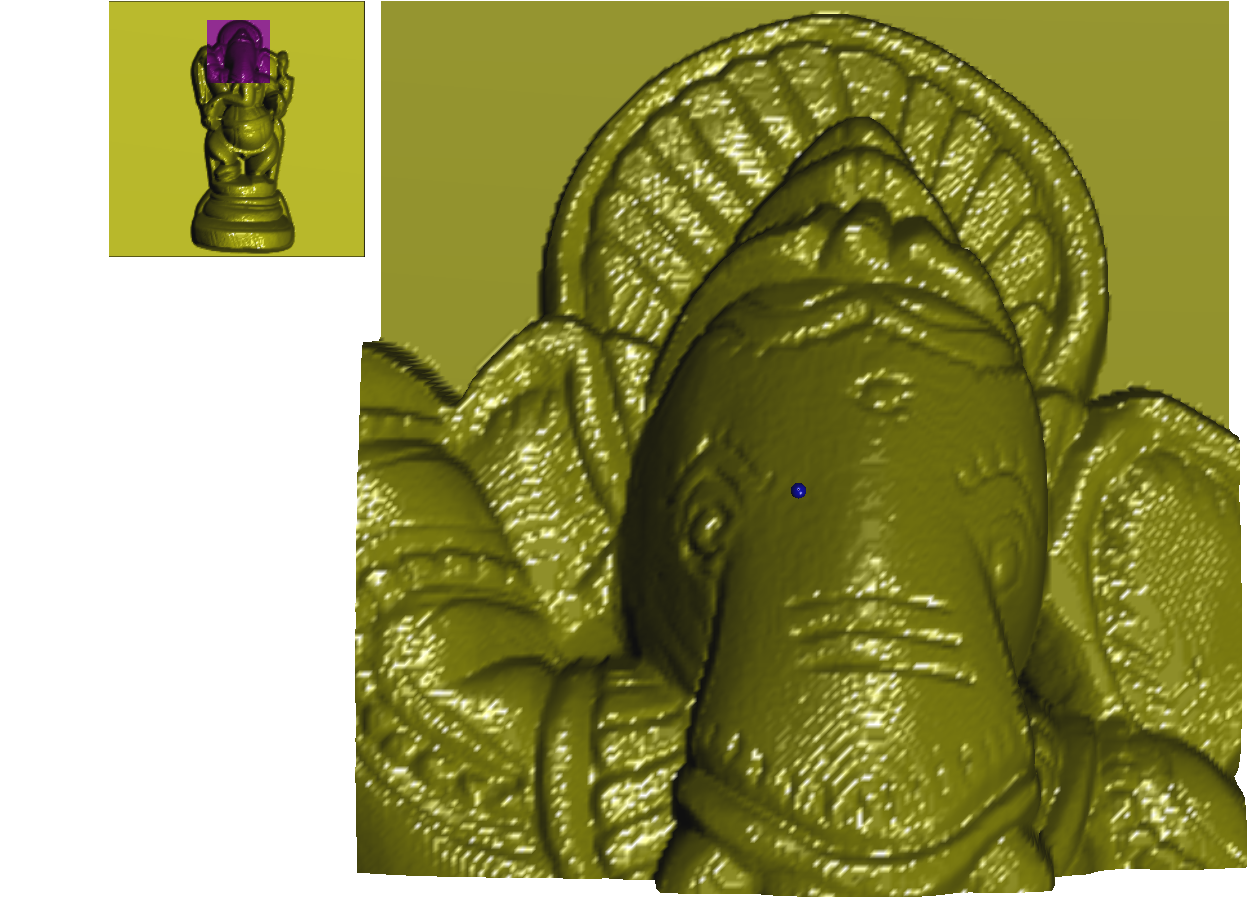}}
  \caption{Model of Ganesh, at (a) least level of details (b) at double the resolution and 
(c) at the finest resolution. (Data Courtesy: \emph{www.archibaseplanet.com})}
  \label{gane_fig}
\end{figure}
\begin{figure} 
  \centering
  \subfloat[]{\label{rack_figa}\includegraphics[width=0.33\textwidth]{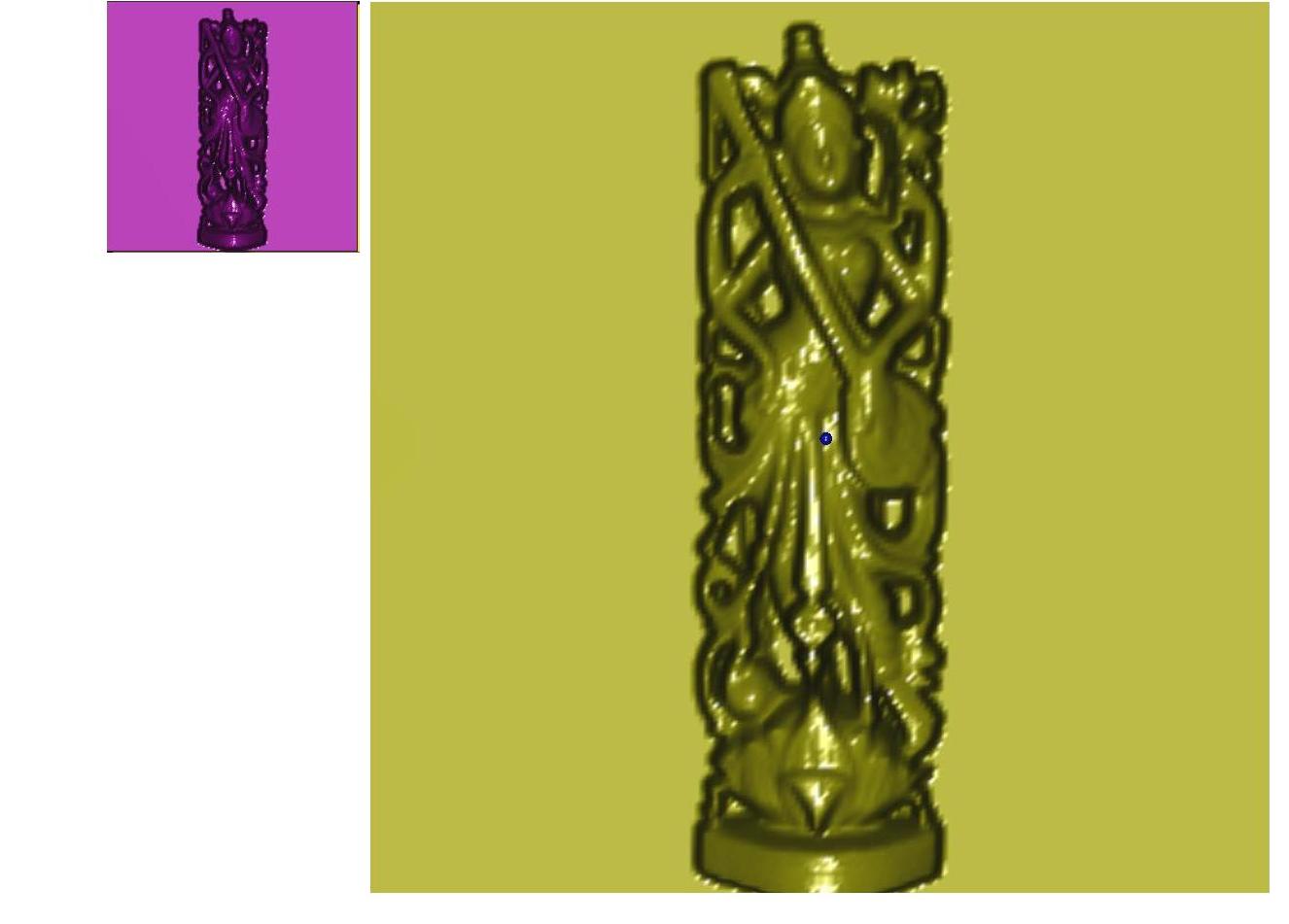}}          
  \subfloat[]{\label{rack_figb}\includegraphics[width=0.33\textwidth]{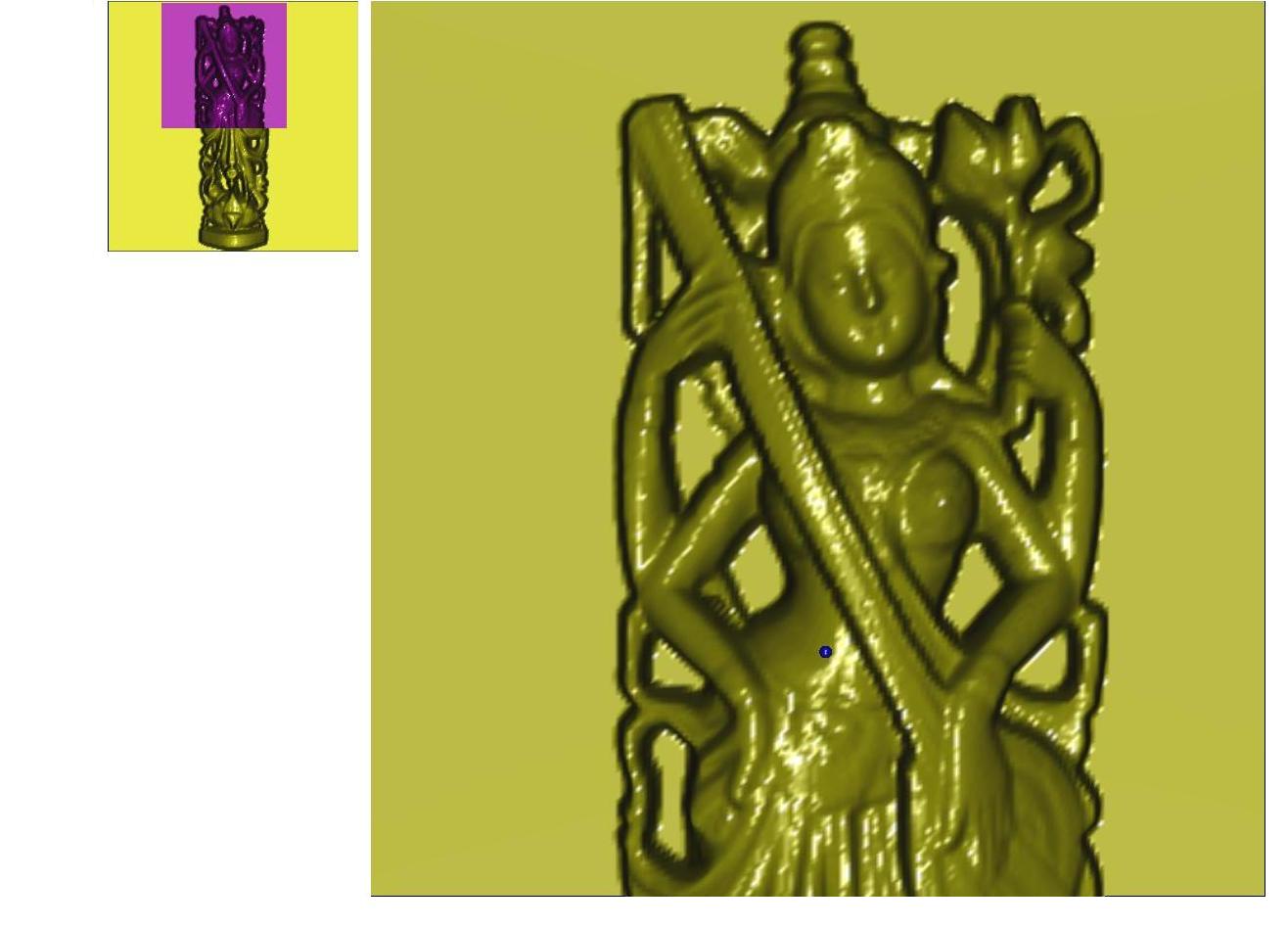}}
  \subfloat[]{\label{rack_figc}\includegraphics[width=0.33\textwidth]{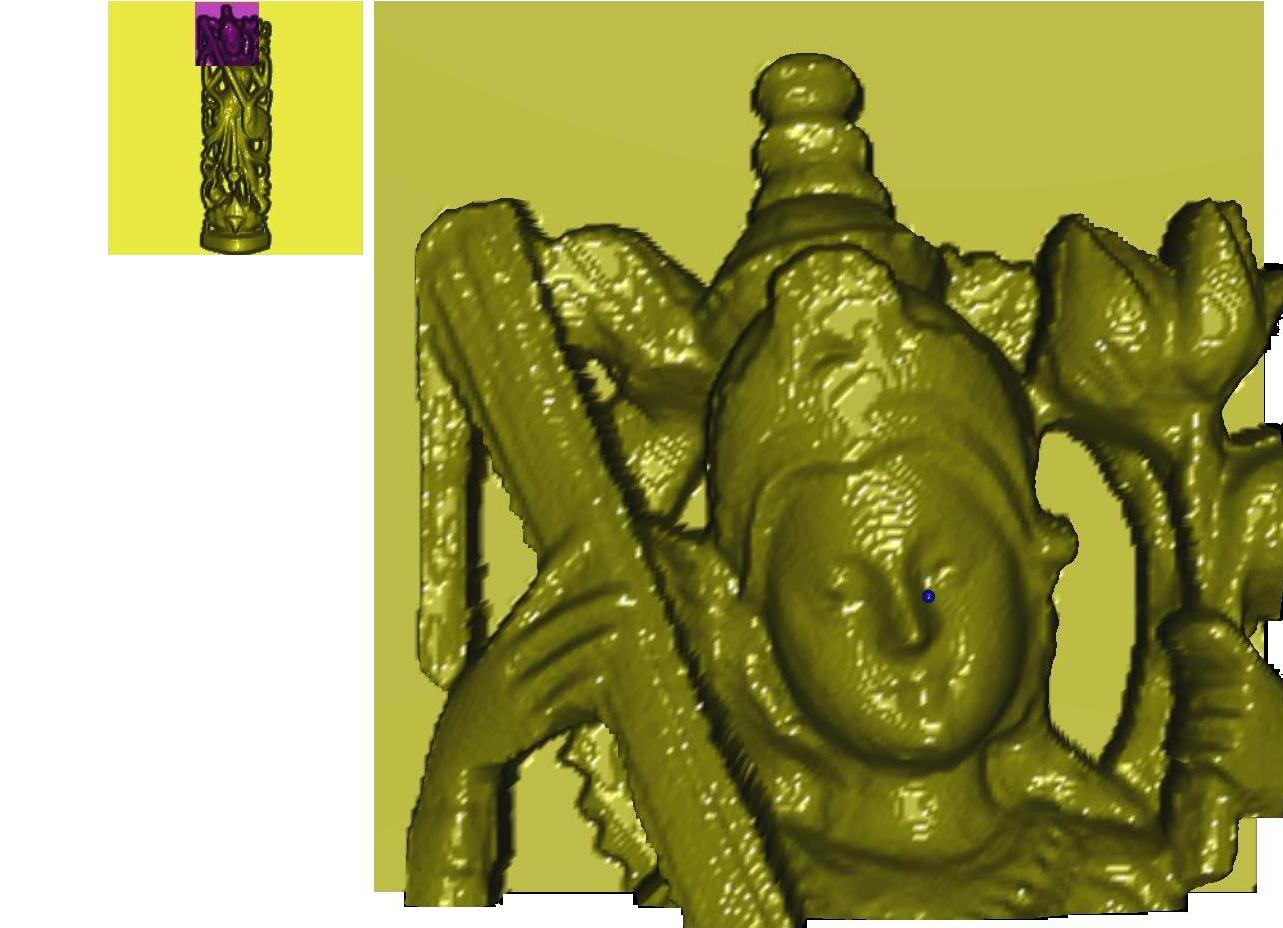}}
 \caption{Model of Goddess Saraswati at (a) level 1 (b) level 2 (c) level 3. (Data Courtesy: \emph{www.archibaseplanet.com}) }
\label{rack_fig} 
\end{figure}

In above cases, each figure consists of two parts where the left part is the reference for 
the users to select the part of the object they wish to explore haptically. The right part of 
the figure corresponds to the selected region at the appropriate resolution for haptic rendering.
Fig. \ref{gane_figc} shows the scaled up version of Fig. \ref{gane_figb}. It is quite clear from 
Fig. \ref{gane_figc} that the users are able to feel even minute details of the sculpture and 
have visual perception of closeness in depth. Hence they can have a more realistic experience. 
The object rendered in Fig. \ref{rack_fig} is a special case that illustrates how one can handle 
holes in the model. For any haptic rendering, holes in the model are difficult to accommodate 
as the proxy would sink through the hole and the user will perceive a wrong depth in the region 
around the holes. We avoid this by defining a base plane on which the object lies. Wherever, 
there is a hole, the depth at that point is replaced by $z(x,y)=z_{MAX}$ where $z_{MAX}$ is the 
maximum depth. This object has several holes, but the users reported a very good experience even in presence 
of such holes. Fig. \ref{rack_figa} allows rendering at a coarser level while Fig. \ref{rack_figb} 
and Fig. \ref{rack_figc} allow rendering at a much finer scale.
\begin{figure} 
  \centering
  \subfloat[]{\label{force_figa}\includegraphics[width=0.5\textwidth]{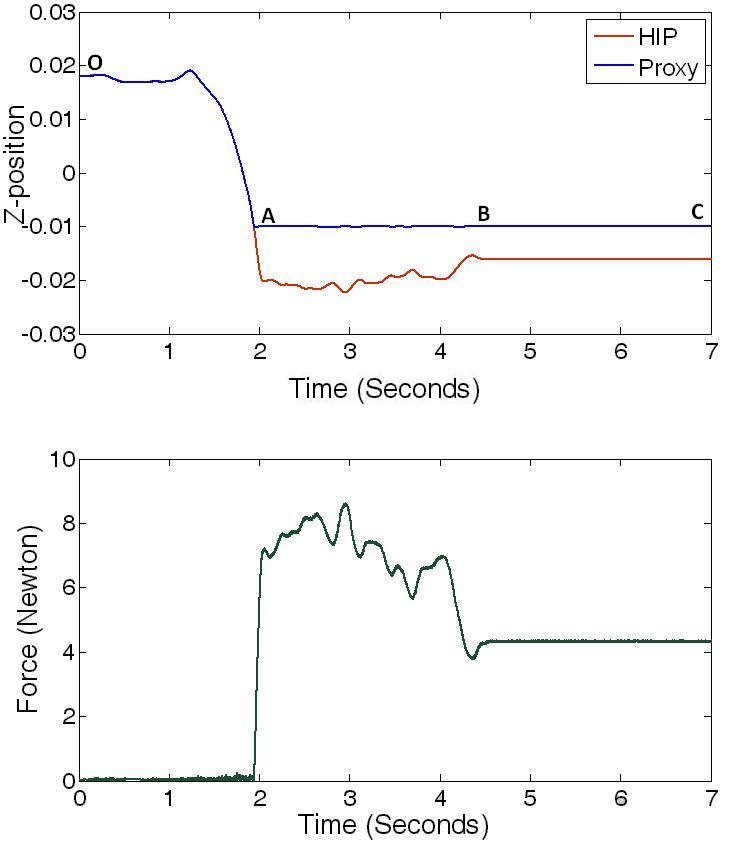}}          
  \subfloat[]{\label{force_figb}\includegraphics[width=0.5\textwidth]{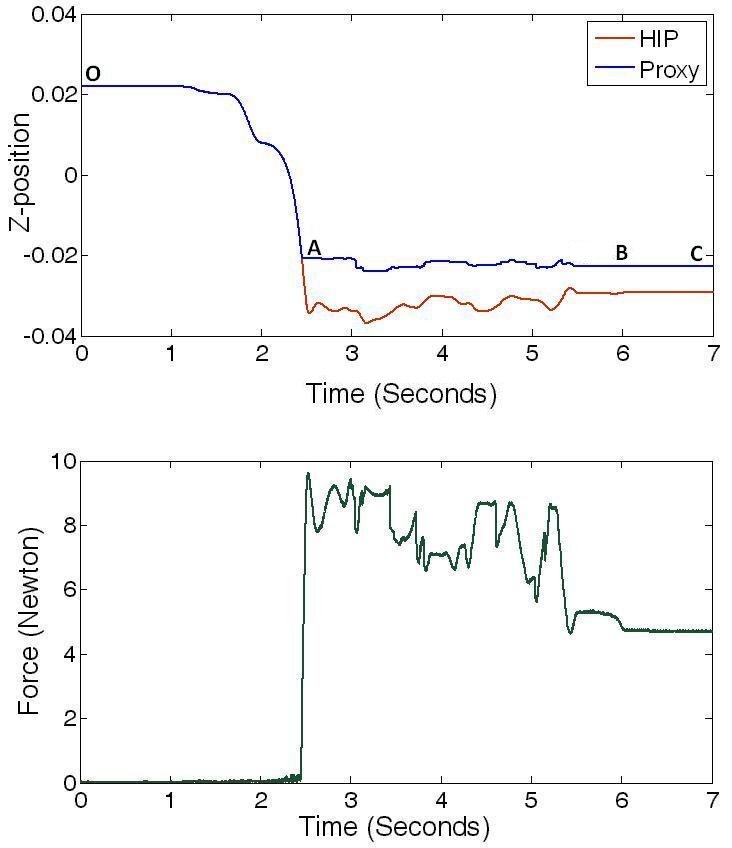}}  
 \caption{Force vs. time graph for a particular interaction with the depth data a) on a flat region b) on a curved region
(The top figure correspond to changes in z-coordinate only).}
 \label{force_fig} 
\end{figure}

Validation of result is often a difficult task during haptic rendering, we demonstrate this using 
Fig. \ref{force_fig} that shows the reaction force versus time relation while haptically interacting with the depth data. The red
line and the blue line in the figure shows the z-component of the HIP and the proxy point,
respectively, during the interaction. The reaction force on the haptic device is also shown during 
the same time interval. In free space the HIP and proxy positions are almost the same as shown in part OA 
of the HIP position and hence the reaction force on the haptic device
is zero. As the HIP penetrates the object the proxy stays on the 
surface according to the iteration method discussed in section \ref{meth_sec}. The proxy point moves continuously
during interaction, whenever there is a change in HIP position. This is shown with the part AB in the curve.
After the point B the HIP position is kept constant inside the objects. As soon as the HIP is kept constant, the proxy 
quickly attains a stable position as shown in the Section BC in the curve. Fig. \ref{force_figa} shows the 
plots corresponding to the interaction on a flat region and Fig. \ref{force_figb} the same on a curved region,
when a larger variation in force is observed.  
\begin{figure} 
\includegraphics[width=0.6\textwidth]{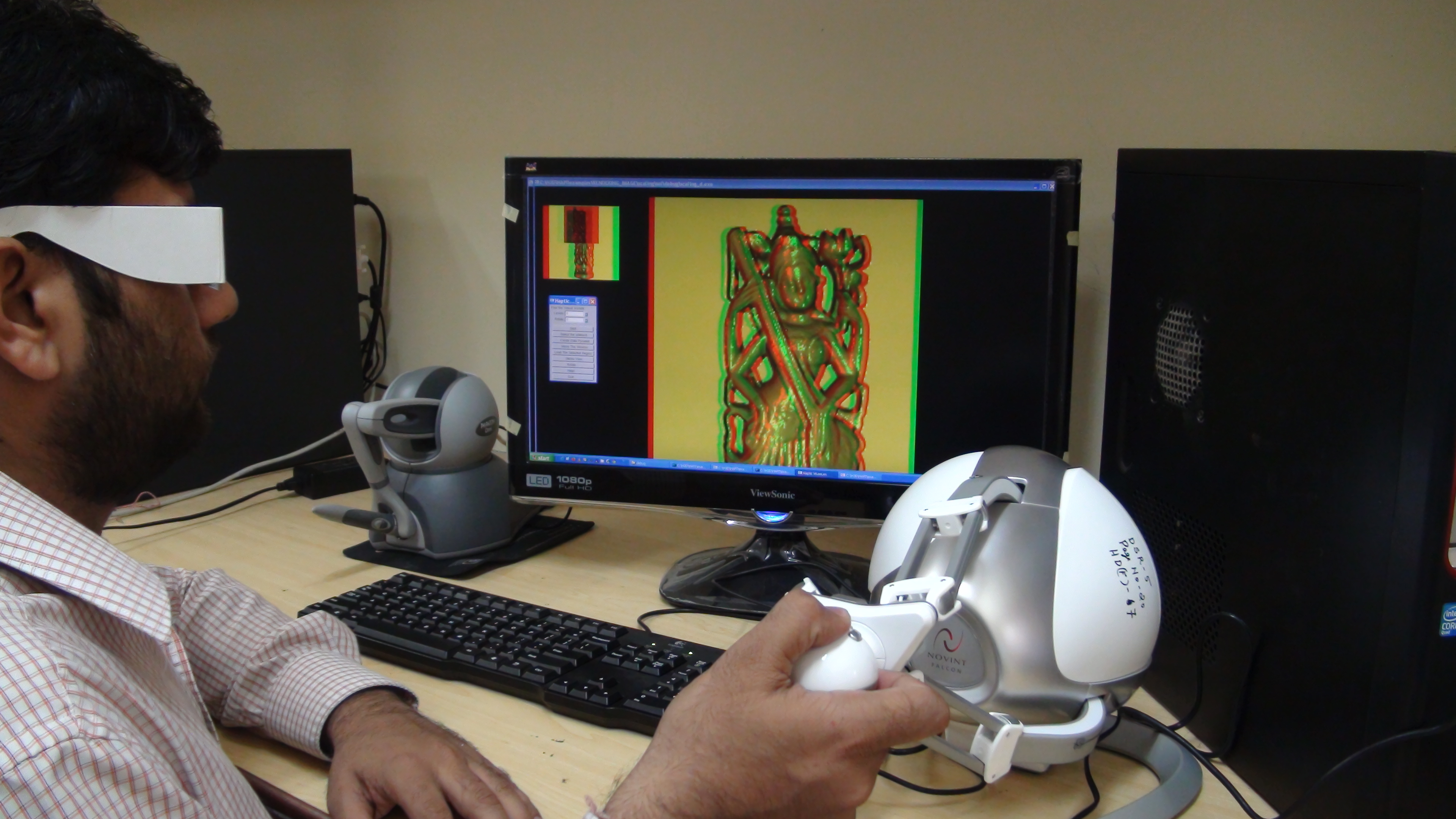}
\centering
\caption{Illustration of hapto-visual immersion of a subject for a virtual cultural heritage 
object. On the right, the user wearing anaglyphic glasses is holding the FALCON haptic device 
while interacting with the cultural heritage model displayed on the screen.}
\label{setup_fig}
\end{figure} 

In Fig. \ref{setup_fig}, we show the actual set up of our virtual haptic museum. A user wearing 
the anaglyphic glasses watches the stereoscopic visual rendering of the artefact and at the same 
time haptically interacts with the object with his hand. This provides an excellent hapto-visual 
immersion of the subject into the virtual object. However, for the visually impaired users, the 
selection of scale and the location for rendering cannot be based on the small navigation window 
on the screen.  For such subjects, we use the buttons available on the haptic device for the user 
to explore the object at different scales and locations.
\section{Conclusions}\label{sec:16}
In this work we have proposed a new technique of rendering cultural heritage objects represented 
as depth map data. Our primary goal is to provide access of cultural heritage objects and sites 
to the visually impaired people. Additionally, our method gives a better immersive experience to 
the sighted persons. We include scalability and stereoscopic display of $3-D$ models as additional 
features to enhance the realism in experience. We conducted experiments with several $3-D$ models 
of cultural significance. We also tested the rendering technique with some subjects and observed that hapto-visual 
rendering of virtual $3-D$ models using the proposed method greatly augmented the user's experience. 
% \subsubsection*{Acknowledgments.} This work was supported in part of a DST grant on Indian digital heritage and another by MCIT on perception engineering.

% order to obtain the discount.
\bibliographystyle{plain}
\bibliography{Thesis_bib}
\end{document}